\documentclass[prl,aps,amsmath,amssymb,longbibliography,superscriptaddress,reprint]{revtex4-1}

\usepackage{graphicx}
\usepackage{dcolumn}
\usepackage{bm}
\usepackage{amsmath}
\usepackage{xcolor}
\usepackage{comment}
\usepackage{siunitx}
\usepackage[colorlinks,linkcolor=blue,urlcolor=blue,citecolor=blue]{hyperref}

\begin{document}

\preprint{APS/123-QED}

\title{Beam Halo Formation via Longitudinal-Transverse Coupling in Continuous-Wave Photoinjectors}%

\author{Zhen Zhang}
\email{zzhang@slac.stanford.edu}
\affiliation{SLAC National Accelerator Laboratory, Menlo Park, California 94025, USA}
\author{Yuantao Ding}
\affiliation{SLAC National Accelerator Laboratory, Menlo Park, California 94025, USA}
\author{David Cesar}
\affiliation{SLAC National Accelerator Laboratory, Menlo Park, California 94025, USA}
\author{Feng Zhou}
\affiliation{SLAC National Accelerator Laboratory, Menlo Park, California 94025, USA}
\author{Ji Qiang}
\affiliation{Lawrence Berkeley National Laboratory, Berkeley, California 94720, USA}
\author{Zhirong Huang}
\affiliation{SLAC National Accelerator Laboratory, Menlo Park, California 94025, USA}

\date{\today}
 
\begin{abstract}
Beam halo formation poses a critical challenge for high-repetition-rate continuous-wave (CW) free-electron lasers (FELs), directly affecting beam quality and machine protection, as observed during the LCLS-II commissioning. We identify and experimentally validate a previously unrecognized three-step mechanism for halo generation in the photoinjector, arising from coupled longitudinal–transverse dynamics in the low-energy beam. Theoretical analysis reveals that (i) the RF buncher induces an energy–radius correlation, (ii) velocity bunching transforms this correlation into hollowed density structures in the bunch head and tail, and (iii) differential overfocusing of these hollowed regions by downstream focusing forms the observed halo. This mechanism is confirmed by particle-in-cell simulations and direct experimental measurements, including controlled formation of a core–ring profile via solenoid tuning. The results establish the physical origin of the halo and demonstrate a mitigation via buncher compression tuning that reduces halo and downstream loss, supporting sustained high-rep-rate FEL operation.
\end{abstract}

\maketitle
Beam halo, defined as a dilute population of particles extending far into the transverse phase space, constitutes a fundamental limitation to accelerator performance. Its presence leads to uncontrolled beam loss, activation of accelerator components, and degradation of the beam core quality, making its control essential for reliable, high-power operation across accelerator complexes\,\cite{allen2002beam,fedotov2006beam}. Beyond its operational consequences, halo formation also reflects fundamental aspects of collective beam dynamics, coupling between degrees of freedom, and nonlinear phase-space evolution in intense charged-particle systems—phenomena of broad relevance across accelerator and beam physics.

Extensive research has elucidated the mechanisms governing hadron beam halo, where core-to-halo transfer is predominantly driven by space-charge-induced parametric resonances and nonlinear collective dynamics\,\cite{gluckstern1994analytic,lagniel1994halo,wangler1998particle,fedotov1999halo,qiang2000beam,qiang2002macroparticle}. Conversely, investigations into electron beam halo have historically centered on circulating machines: Energy Recovery Linacs (ERLs) and storage rings. In ERLs, halo generation is linked to space-charge mismatch and nonlinear radio-frequency (RF) forces that can lead to contamination or quenching of superconducting RF cavities\,\cite{fedotov2003mechanisms,tanaka2018new,tanaka2020beam}. In storage rings, long-term diffusion processes—stemming from nonlinear resonances, intrabeam scattering, and beam–gas collisions—produce halo particles that ultimately restrict dynamic aperture and photon beam stability\,\cite{wang2017beam,yang2018evaluation,hwang2021beam,yang2025beam}.

Halo formation in single-pass electron linacs, particularly in the low-energy photoinjector stage, has received comparatively less attention. This is largely because historical, pulsed normal-conducting linacs operated at low average power and low repetition rate (see, for example, Ref.\,\cite{akre2008commissioning}). However, the emergence of continuous-wave (CW) superconducting free-electron laser (FEL) facilities (e.g., LCLS-II\,\cite{raubenheimer2015lcls}, SHINE\,\cite{zhu2017sclf} and European XFEL upgrade\,\cite{kostin2019progress}) marks a fundamental shift in requirements. Operating at high average current with extremely stringent loss tolerances, these systems are acutely sensitive to halo generated early in the injector. Such halo populations directly degrade beam brightness, significantly complicate lattice matching, and accumulate unacceptable losses in the downstream beamline that can interrupt continuous high-repetition-rate operation. A detailed understanding of the dominant mechanism responsible for halo generation in CW photoinjectors is therefore critical for achieving stable, low-loss FEL operation.

The present study is motivated by experimental observations during the commissioning of the CW photoinjector at the LCLS-II. Distinct halo structures were routinely observed in transverse beam images downstream of the injector. The beam profile exhibited pronounced non-Gaussian features—ranging from diffuse outer rings to characteristic donut shapes—indicating complex beam dynamics not captured by conventional injector models. While this study focuses on the LCLS-II photoinjector, the mechanism identified here is generically applicable to most CW photoinjectors employing velocity bunching at sub-MeV energies—a configuration increasingly adopted in superconducting-based FELs\,\cite{zhou2017lcls,jiang2024design,bazyl2021cw} and ERLs\,\cite{chen2025design}.

\begin{figure*}[htb]
   \centering
   \includegraphics[width=0.9\textwidth]{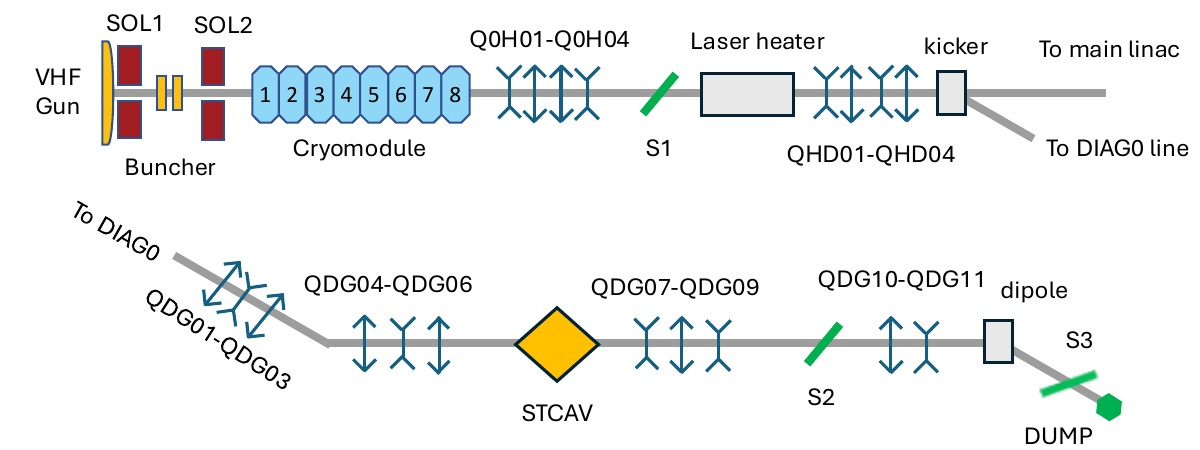}
   \caption{Layout of the LCLS-II photoinjector showing the VHF gun, buncher, solenoids (SOL1 and SOL2), superconducting cryomodule with eight cavities, quadrupoles (elements labeled “Q”), and diagnostic beamline with STCAV and imaging screens (S1-S3) used for halo characterization.}
   \label{fig:layout}
\end{figure*}

Figure~\ref{fig:layout} shows the schematic layout of the LCLS-II photoinjector studied in this work\,\cite{zhou2017lcls}. The electron beam at a nominal bunch charge of \qty{70}{pC} is generated in a very-high-frequency (VHF) normal-conducting RF gun\,\cite{sannibale2012advanced} and immediately focused by the first solenoid (SOL1) for emittance compensation\,\cite{carlsten1989new,serafini1997envelope}. The beam energy after the gun is approximately \qty{650}{keV}. A normal-conducting buncher cavity imparts an energy chirp that enables velocity bunching. The second solenoid (SOL2) provides additional transverse focusing and matches the beam into a superconducting cryomodule, which accelerates it to about \qty{75}{MeV} before the laser heater. All  cavities in the cryomodule are operated on-crest, except for the second cavity, which is either turned off or set to the zero-crossing phase to provide additional compression following the buncher. The resulting beam maintained an rms bunch length of approximately \qty{0.9}{mm}. Downstream of the laser heater, a fast kicker directs the beam either toward the main linac or into a dedicated diagnostic beamline equipped with an S-band transverse deflecting cavity (STCAV) for time-resolved profile measurements. Quadrupoles are grouped to control beam size, perform emittance measurements, and enable matching. Three screens in the layout are specifically used for beam halo measurements: one after the cryomodule (S1), one after the STCAV (S2), and one after the bending magnets before the dump (S3). This configuration allows detailed experimental validation of the theoretical model and simulations presented in this work.

In this Letter, we propose and validate a three-step mechanism responsible for the observed halo formation in the CW photoinjector. The process originates from the coupling between longitudinal and transverse dynamics at low beam energy. Before acceleration in the cryomodule, the beam energy remains below \qty{1}{MeV}, where even small differences in particle kinetic energy lead to appreciable variations in drift velocity. Particles with higher longitudinal momentum move faster and can overtake slower ones, producing the familiar velocity-bunching behavior\,\cite{serafini2001velocity,anderson2005velocity}.

The beam dynamics of this sub-MeV electron beam are primarily governed by the combined effects of the buncher’s RF field and the internal space-charge forces. In the buncher, the longitudinal electric field follows\,\cite{wangler2008rf}
\begin{align}
    E_z(r)=E_z(0)J_0(k_rr)\approx E_z(0)\left(1-\frac{k_r^2r^2}{4}\right)\,,
\end{align}
where $k_r$ is the transverse wave number of the RF mode. The field amplitude decreases quadratically with radius. The space-charge field exhibits a similar pattern: it accelerates on-axis electrons more strongly in the bunch head and decelerates them more strongly in the tail than for off-axis particles\,\cite{reiser2008theory,wu2008analytical}. Together, these effects establish a quadratic correlation between longitudinal momentum and radius, which can be expressed as
\begin{align}
    \delta = (h_1z+h_2z^2)\left(1-k^2r^2\right)\,,
\end{align}
where $\delta$ is the relative energy deviation, $z$ is the longitudinal coordinate within the bunch (with $z=0$ at the bunch center), $h_1$ and $h_2$ are the first- and second-order energy chirp coefficients, and $k$ represents the effective strength of the radial dependence, which varies along the bunch. This constitutes the first step of the proposed mechanism, wherein the combined RF and space-charge fields imprint a transverse dependence on the longitudinal momentum, establishing a correlation between $p_z$ and $r$ within each longitudinal slice.

During subsequent velocity bunching, this $r$-dependent energy chirp causes a curved compression surface in $(r,z)$ space for each slice. This differential compression leads naturally to the formation of hollowed regions at the beam head and tail—the second stage in our proposed three-step mechanism. The longitudinal coordinate $z_f$ after the velocity bunching can then be mapped as
\begin{align}
    z_f = z+R_{56}\delta+T_{566}\delta ^2\,,
\end{align}
where $R_{56}$ and $T_{566}$ are the first- and second-order longitudinal momentum compaction coefficients, respectively. In our sign convention ($z$ increasing toward the bunch head), for a drift of length $L$, $R_{56}\approx L/\beta\gamma^2$ and $T_{566} \approx -(L/\beta)(3\gamma^2-1)/(2\gamma^4)$, describing the dependence of the path length (or arrival time) on particle energy\,\cite{minty1999beam}.

To model the halo formation process, we assume the beam has a football-shaped geometry with a longitudinally varying transverse size
\begin{align}
    R(z) = R_{max}\sqrt{1-(z/Z_0)^2}\,,\quad -Z_0\leq z \leq Z_0
\end{align}
where $Z_0$ defines the half-length of the bunch. Assuming the beam is uniformly filled, the initial density within each longitudinal slice is
\begin{align}
    n_0(r;z)=\rho r\,,\quad 0\leq r \leq R(z)
\end{align}
where $\rho$ is a normalization constant ensuring that integration over the full volume yields the total bunch charge. The linear dependence on $r$ reflects that $n_0(r;z)$ represents the area density integrated over a circular ring of radius $r$ within each slice.
For a given final slice $[z_1,z_2]$ after velocity bunching, the set of initial $z$ that maps into the final slice can be expressed as
\begin{align}
    I_r = \{\, z \in [-Z_0, Z_0]:\, z_1\leq F(z;r)\leq z_2,\, r\leq R(z)\}
\end{align}
where 
\begin{align}
    F(z;r) = a_rz^2+b_rz\,,
\end{align}
with
\begin{align}
    a_r &= R_{56}h_2(1-k^2r^2)+T_{566}h_1^2(1-k^2r^2)^2\,,\\
    b_r &= 1+R_{56}h_1(1-k^2r^2)\,.
\end{align}
$I_r$ can be written as the union of intervals $\cup_j[A_j,B_j]$, where each $A_j$ and $B_j$ is a root of $F(z;r)=z_1$ or $F(z;r)=z_2$. The average radial density $\bar{n}(r)$ for the slice $[z_1,z_2]$ after velocity bunching can be expressed as
\begin{align}
    \bar{n}(r;z_1,z_2) = \rho r\frac{\sum_j(B_j-A_j)}{|z_2-z_1|}\,.
\end{align}

\begin{figure}[htb]
   \centering
   \includegraphics[width=0.95\columnwidth]{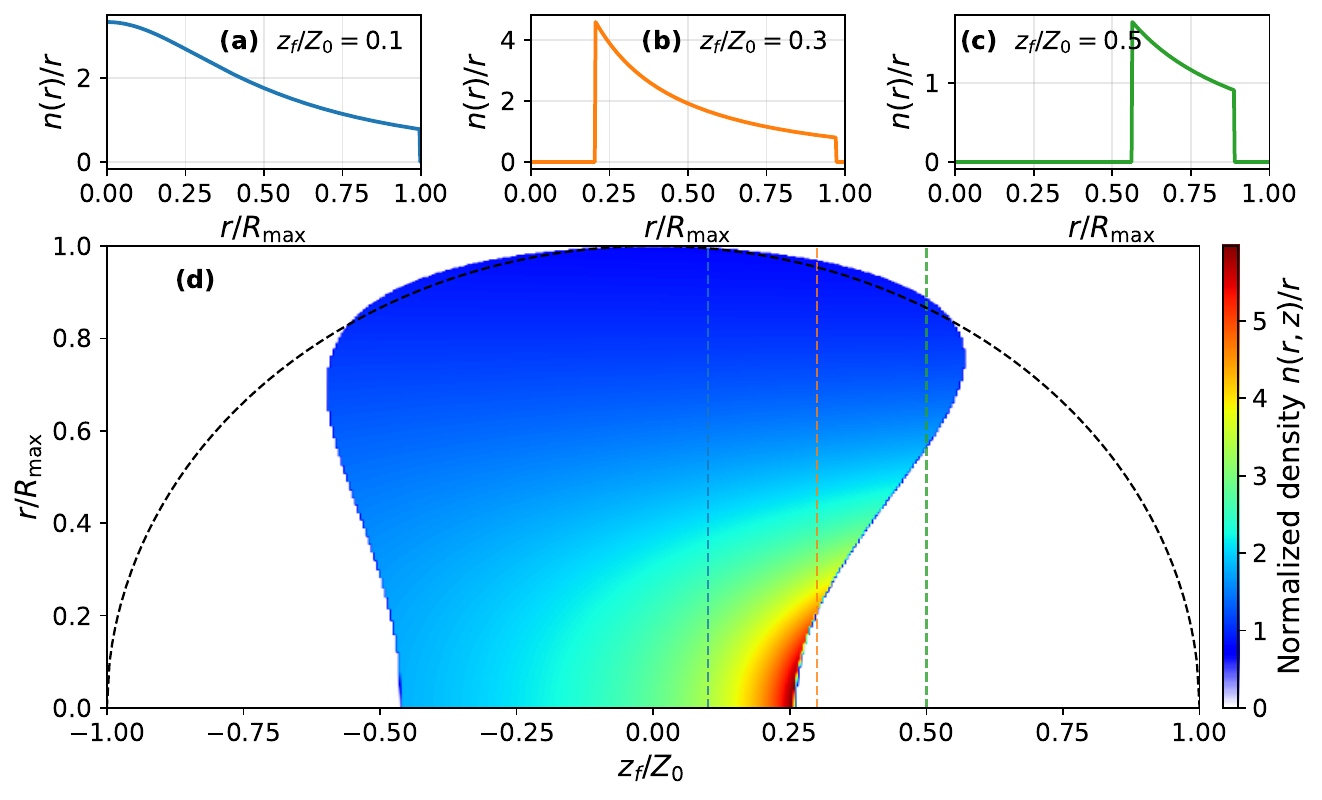}
   \caption{Numerical solution of the analytical model showing evolution from an initially uniform beam to hollowed head- and tail-regions during velocity bunching. (a)–(c) Radial density profiles $n(r)/r$ at selected normalized positions $z_f/Z_0$ shown by the colored dashed lines in (d). (d) Two-dimensional density $n(r,z)/r$, normalized to the initial uniform density. The dashed curve indicates the initial boundary. The beam head is on the right.
}
   \label{fig:model}
\end{figure}

This formulation enables direct numerical evaluation of the post-compression density distribution and provides quantitative predictions of how the initially uniform beam develops hollowed structures through $r$-dependent velocity bunching. Figure\,\ref{fig:model} shows that an initially uniform, elliptically bounded beam evolves into a structure with hollowed regions at both the head and tail, where on-axis particles move faster toward the beam center. The density $n(r)/r$ is shown to eliminate the geometrical factor of a circular ring at a radius $r$. The parameters used in the calculation are $R_{max}=$\qty{10}{mm}, $Z_0=$\qty{2}{mm}, $\gamma=2.5$, $L=$\qty{1.8}{m}, $h_1 =-2~\mathrm{m^{-1}}$, $h_2 =-160~\mathrm{m^{-2}}$, and $k = 120~\mathrm{m^{-1}}$, which are similar with the beam parameters after the buncher in the LCLS-II injector. These hollowed regions originate from slices where the $r$-dependent velocity chirp generates the strongest curvature in the $(r,z)$ compression mapping. The asymmetry between the head and tail arises from the nonlinear terms in the energy chirp and longitudinal momentum compaction. It is worth noting that because $r$ is fixed at this stage, the projected transverse profile remains unchanged, despite the emergence of hollowed regions in the three-dimensional phase space.

A relatively strong transverse correlation parameter $k$ is chosen to clearly illustrate the formation of the hollowed structure, as the simplified model neglects space-charge effects. In reality, space-charge forces significantly enhance this process. Off-axis particles that advance to the head or lag at the tail experience additional push from their self-fields, further amplifying their separation. At the same time, the space-charge fields partially counteract the RF-induced chirp, leading to nonlinear longitudinal compression where compression and de-chirping occur simultaneously. Consequently, the parameters of the space-charge-free model are adjusted to reproduce the qualitative behavior of a realistic beam, demonstrating that this simplified analytical model captures the essential mechanism behind the formation of hollowed structures during velocity bunching.

The formation of the transverse halo represents the final stage of this process. In the hollowed regions generated during velocity bunching, particles experience substantially weaker transverse space-charge forces than those in the dense core. This is a geometric consequence of the ring-like distribution: electrons near the inner edge feel little net radial repulsion because most charge lies at larger radii, while particles near the outer edge experience reduced self-forces simply because the hollowed structure encloses far less charge than a dense core at the same radius. In contrast, all particles in the hollowed region reside at larger transverse radii and therefore experience stronger external focusing from the solenoids, quadrupoles, and RF cavities. Under these focusing forces, the weakly repelled hollow-region electrons are readily over-focused and pushed to even larger radii, forming a diffuse, extended population around the beam core. This mechanism naturally converts the hollowed structures formed during velocity bunching into the transverse halo observed downstream.

\begin{figure}[htb]
   \centering
   
   \includegraphics[width=0.9\columnwidth]{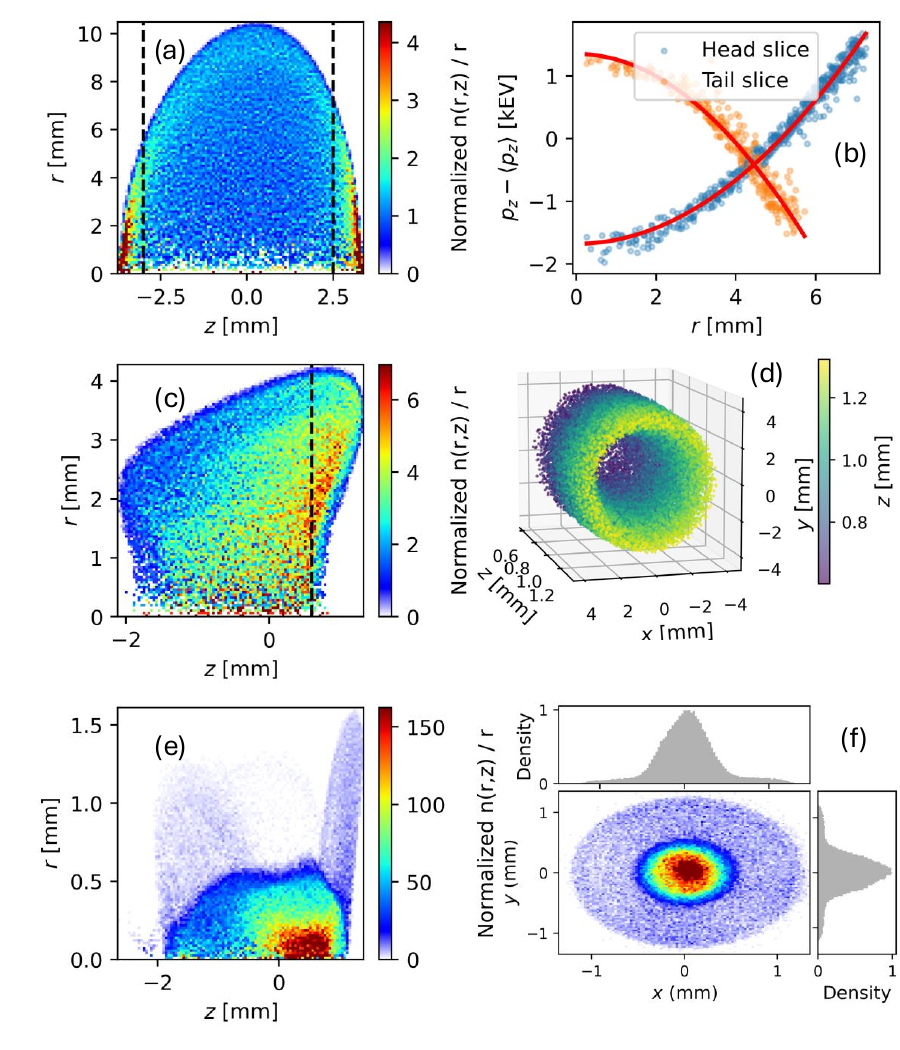}
   \caption{Simulation of beam halo formation in the LCLS-II photoinjector.
(a) Two-dimensional density $n(r,z)/r$ after the buncher. (b) Correlation between longitudinal momentum $p_z$ and radius $r$ for representative head and tail slices.
(c) $n(r,z)/r$ distribution at the cryomodule entrance, showing hollowed regions at the beam head and tail.
(d) Three-dimensional view of the beam head, right portion of dashed line in (c).
(e) $n(r,z)/r$ distribution in the middle of cryomodule.
(f) Transverse beam profile after the cryomodule, showing a bright core surrounded by a diffuse halo. The beam head is on the right in (a), (c), and (e).
}
\label{fig:simulations}
\end{figure} 

Building upon the analytical model, we performed start-to-end particle-in-cell simulations with Astra\,\cite{flottmannAstra} and experimental measurements to validate the proposed mechanism of beam halo formation. In the simulation, the beam charge was \qty{70}{pC}, and the buncher cavity provided the primary beam compression while the second cavity in the cryomodule was turned off\,\cite{zhang2025fault}. The simulation results in Fig.\,\ref{fig:simulations} validate the three-step mechanism described above. As shown in Figs.\,\ref{fig:simulations}(a) and \ref{fig:simulations}(b), the beam exhibits a pronounced correlation between longitudinal momentum and transverse radius immediately after the buncher. In the head region, off-axis particles gain higher momentum, while in the tail they lose it, in excellent agreement with the analytical prediction. As the beam propagates toward the cryomodule, the curvature in the longitudinal–transverse phase space becomes increasingly nonlinear, giving rise to hollowed density structures at both ends of the bunch, as seen in Figs.\,\ref{fig:simulations}(c) and \ref{fig:simulations}(d). These hollowed region particles are subsequently overfocused to large radii, forming a diffuse outer population surrounding the central core (Fig.\,\ref{fig:simulations}(e)). The final transverse beam profile (Fig.\,\ref{fig:simulations}(f)) displays a bright, compact core enveloped by an extended halo, closely resembling experimental observations during injector operation. These simulations thus confirm the complete physical picture: an r-dependent energy chirp generated by the buncher, the emergence of hollowed structures during velocity bunching, and their transformation into a beam halo through differential focusing downstream.

\begin{figure}[htb]
   \centering
   \includegraphics[width=0.95\columnwidth]{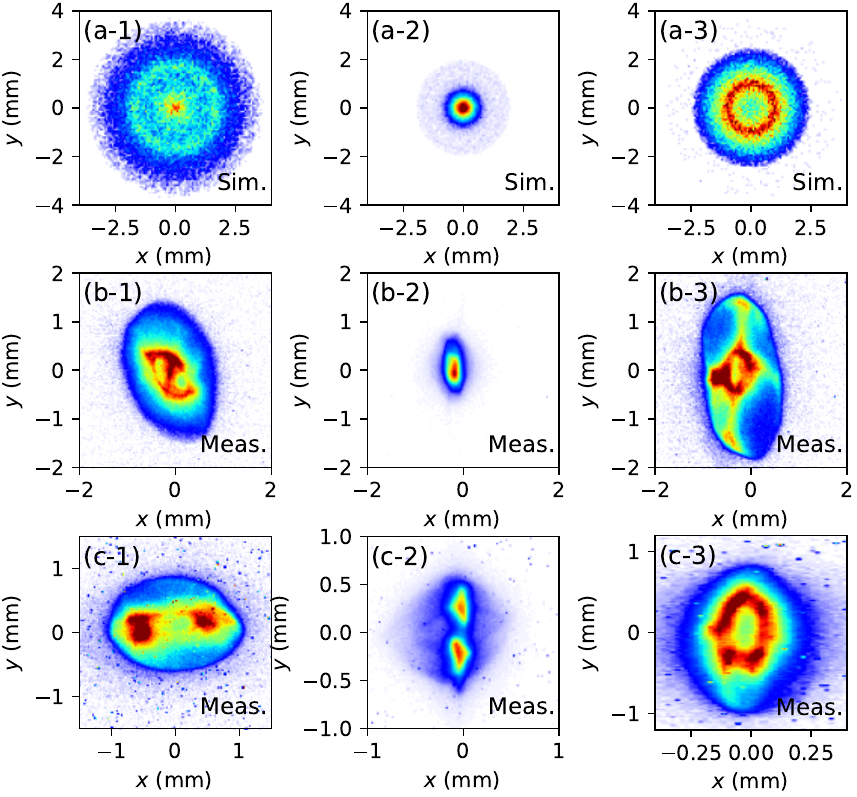}
   \caption{Simulated and measured beam profiles at screen S1 showing core–halo focusing. (a-1)–(a-3) Simulations for increasing SOL2 with upstream quadrupoles off. (b-1)–(b-3) Measurements under similar conditions. (c-1)–(c-3) Profiles with quadrupoles on, reshaping the beam into horizontal lobes, vertical lobes, or a circular ring. Colormaps match Fig.\,\ref{fig:model}(d), normalized from zero to each image’s maximum.}
\label{fig:scan_SOL2}
\end{figure} 

Due to the limited diagnostics available in the injector, direct observation of the halo formation process predicted by the model is not feasible. Beam images are obtainable only at discrete screens, making it difficult to track the evolution of hollowed structures along the beamline. Instead, we designed a simplified experiment to probe the characteristic focusing behavior of a core–halo beam and compare it with simulations. The upstream matching quadrupoles (Q0H01–Q0H04) (as shown in Fig.~\,\ref{fig:layout}) were intentionally turned off, and the strength of the second solenoid (SOL2) was gradually increased to monitor the focusing response at screen S1. The buncher phase was set to \qty{-60}{\degree} from on-crest, corresponding to strong compression. Figure~\ref{fig:scan_SOL2} summarizes the results: the first row (a-1)–(a-3) shows simulations, while the second row (b-1)–(b-3) presents corresponding measurements under similar conditions. As SOL2 increases, both simulation and experiment display the same qualitative trend—the beam evolves from a ring with a bright core to a compact spot and finally to an overfocused ring—consistent with the predicted difference in focusing response between the core and halo. When the quadrupoles are reactivated, the strong halo component enables deliberate reshaping of the beam into two horizontal, two vertical, or circular configurations (Figs.\,\ref{fig:scan_SOL2}(c-1)–(c-3)). These complex profiles, first observed during LCLS-II injector commissioning and later reproduced in dedicated experiments, can arise even in the absence of laser, cathode, or field imperfections, indicating that they originate intrinsically from core–halo dynamics. Their presence complicates accurate beam characterization and can lead to beam loss downstream, underscoring the importance of mitigating halo formation for stable CW FEL operation.

\begin{figure}[htb]
   \centering
   \includegraphics[width=0.95\columnwidth]{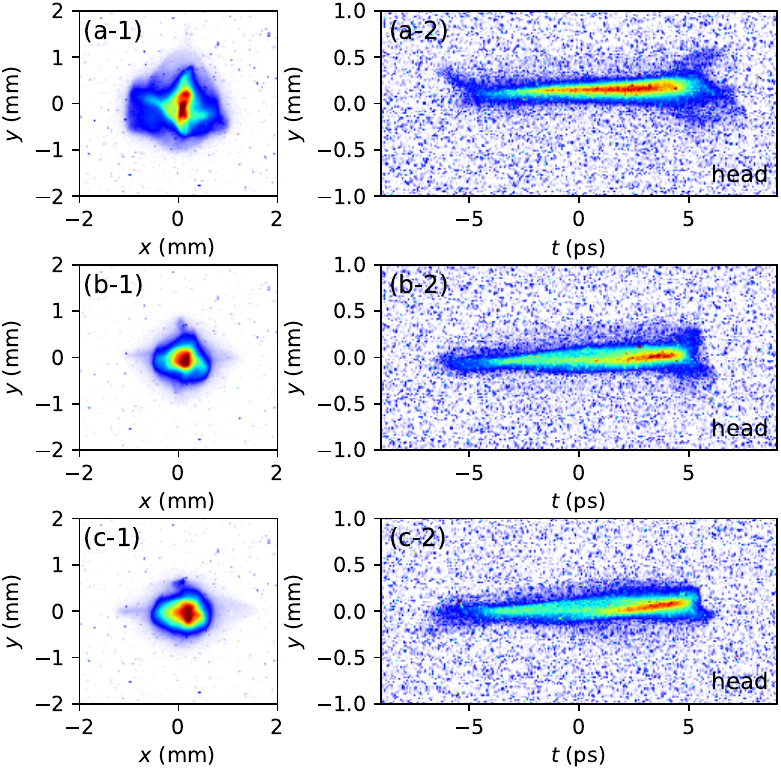}
   \caption{Measured profiles for three buncher phases showing halo mitigation. (a-1)–(c-1) Transverse $x-y$ profiles at screen S1 before the laser heater. (a-2)–(c-2) Time-resolved $y-t$ images at screen S2 after STCAV. The beam head is on the right. The buncher phases are (a) \qty{-55}{\degree}, (b) \qty{-46}{\degree}, and (c) \qty{-40}{\degree} from on-crest. Colormaps match Fig.\,\ref{fig:model}(d), normalized from zero to each image’s maximum.}
\label{fig:STCAV}
\end{figure} 

Another set of measurements focused on the time-resolved characterization of halo particles, which primarily originate from the beam head and tail. As predicted by the three-step model, the most direct approach to mitigating beam halo formation is to reduce velocity bunching at low energy—namely, to weaken the compression applied by the buncher cavity. This limits the depth of the hollowed structures that seed the halo. We experimentally tested this strategy at the LCLS-II by varying the buncher phase. To maintain a similar bunch length at the injector exit, the second cavity in the cryomodule was operated at the zero-crossing phase to provide modest velocity bunching at higher energy (around \qty{7}{MeV}), where the reduced beam size and weaker space-charge forces greatly suppress the formation of hollowed structures. For each setting, the residual energy chirp was minimized at screen S3, and the beam optics were rematched to maintain the target Twiss parameters at screens S1 and S2. Figure\,\ref{fig:STCAV} presents the measured transverse profiles at screen S1 and time-resolved streak images at screen S2 obtained using STCAV. For strong compression at a buncher phase of \qty{-55}{\degree} from on-crest, the transverse image before the laser heater (Fig.\,\ref{fig:STCAV}(a-1)) exhibits a diffuse halo surrounding the bright core. The corresponding time-resolved image (Fig.\,\ref{fig:STCAV}(a-2)) shows pronounced head- and tail-side expansions consistent with the predicted hollowed structures. As the buncher phase increases to \qty{-46}{\degree} and \qty{-40}{\degree}, the halo features at both S1 and S2 (Figs.\,\ref{fig:STCAV}(b) and \ref{fig:STCAV}(c)) progressively diminish, confirming that reduced velocity bunching effectively suppresses halo formation. This operational strategy has since been implemented in the LCLS-II, leading to a marked reduction in beam halo and associated downstream beam losses.

In conclusion, we have identified and experimentally confirmed a three-step mechanism responsible for beam halo formation in a CW photoinjector. The process begins with an $r$-dependent energy chirp induced by the RF buncher, followed by hollowed structure formation during velocity bunching, and ends with overfocusing of these regions into a halo by downstream optics. Particle-in-cell simulations and experimental measurements in the LCLS-II injector confirm this mechanism and demonstrate that halo formation can be effectively mitigated by reducing velocity bunching at low beam energy. Although demonstrated at the LCLS-II, the underlying mechanism is generic to photoinjectors employing low-energy compression, a configuration shared by nearly all CW superconducting FEL facilities under development worldwide. This understanding provides a practical pathway toward stable, low-loss operation of high-brightness CW FELs and offers a general framework for addressing halo control in future photoinjector designs.

The authors thank D. Dowell, T. Vecchione and J. Tang for their assistance and helpful discussions. This work was supported by the U.S. Department of Energy under Contract No. DE-AC02-76SF00515.

\bibliographystyle{IEEEtran}
\bibliography{ref}

\end{document}